\documentclass{iau}
\usepackage{graphicx,natbib}
 
\newcommand{\apj}{ApJ}           
\newcommand{\apjl}{ApJ}           
\newcommand{\mnras}{MNRAS}       

\newcommand{\aap}{A\&A}

\newcommand{\pasp}{PASP}

\title[Counter-rotating disks in galaxies]{Counter-rotating disks in galaxies: dissecting kinematics and stellar populations with 3D spectroscopy}

\author[Coccato et al.]{
Lodovico Coccato$^1$, Lorenzo Morelli$^{2,3}$, Alessandro Pizzella$^{2,3}$, Enrico~Maria
Corsini$^{2,3}$, Elena Dalla Bont\`a$^{2,3}$, Maximilian Fabricius$^{4,5}$
}
\affiliation{$^1$European Southern Observatory, Karl-Schwarzschild-Stra$\beta$e 2, D-85748, Garching bei M\"unchen, Germany.\\
$^2$Dipartimento di Fisica ed Astronomia ``Galileo Galilei'', Universit\`a di Padova, vicolo dell'Osservatorio 3, 35122 Padova, Italy.\\
$^3$INAF-Osservatorio Astronomico di Padova, vicolo dell'Osservatorio 5, 35122 Padova, Italy.\\
$^4$Max Planck Institute for Extraterrestrial Physics, Giessenbachstra$\beta$e, D-85748 Garching, Germany.\\
$^5$University Observatory Munich, Scheinerstra$\beta$e 1, 81679 Munich, Germany.\\ 
\medskip
email: {\tt lcoccato@eso.org} \\}

\pubyear{2014}
\volume{309}
\jname{Galaxies in 3D across the Universe}
\editors{B. L. Ziegler, F. Combes, H. Dannerbauer, M. Verdugo, eds.}

\begin{document}

\maketitle

\begin{abstract}
  We present a spectral decomposition technique that separates the
  contribution of different kinematic components in galaxies from the
  observed spectrum. This allows to study the kinematics and
  properties of the stellar populations of the individual components
  (e.g., bulge, disk, counter-rotating cores, orthogonal
  structures). Here, we discuss the results of this technique for
  galaxies that host counter-rotating stellar disks of comparable
  size. In all the studied cases, the counter-rotating stellar disk is
  the less massive, the youngest and has different chemical content
  (metallicity and $\alpha$-elements abundance ratio) than the main
  galaxy disk. Further applications of the spectral decomposition
  technique are also discussed.

\end{abstract}

\firstsection
\section{Introduction}

Galaxies can have a complex kinematic structure, due to the
superimposition of multiple components that have different kinematics
and stellar populations, such as bulge and disk (in lenticulars and
spirals), host spheroid and polar ring (in polar ring galaxies), and
counter-rotating stellar disks or kinematically decoupled cored
(mainly - but not only - in early-type galaxies).

The interesting case of large-scale counter-rotating galaxies, i.e.,
those that host two stellar components of comparable sizes that rotate
along opposite directions, is the topic of this work. The prototype of
this class of objects is the E7/S0 galaxy NGC 4550, whose
counter-rotating nature was first discovered by \citet{Rubin+92}.

\section{Spectral decomposition}

The spectral decomposition technique by \citet{Coccato+11} is an
implementation of the penalized pixel fitting code by
\citet{Cappellari+04}. For a given spectrum, the code builds two
synthetic stellar templates (one for each stellar component) as a
linear combination of stellar spectra from an input stellar
library. It then convolves these two stellar templates with two
Gaussian line-of-sight velocity distributions (LOSVDs), which are
characterized by different velocity and velocity dispersion. The input
galaxy and stellar spectra are normalized to their continuum level,
therefore the contribution of each component to the observed galaxy
spectrum is measured in terms of light. Gaussian functions are added
to the convolved synthetic templates to account for ionized-gas
emission lines. Multiplicative Legendre polynomials are included in
the fit to match the shape of the galaxy continuum, and are set to be
the same for the two synthetic templates. This accounts for the
effects of dust extinction and variations in the instrument
transmission. These steps are embedded in a iterative $\chi^2$
minimization loop.

The spectral decomposition code returns the spectra of two
best-fit synthetic stellar templates and ionized-gas emissions, along
with the best-fitting parameters of light contribution, velocity, and
velocity dispersion. The line strength of the Lick indices of the two
counter-rotating components are then extracted from the two best-fit
synthetic templates, and used to infer the properties of the stellar
populations of the two components.

\begin{figure}
\includegraphics[width=1.0\columnwidth]{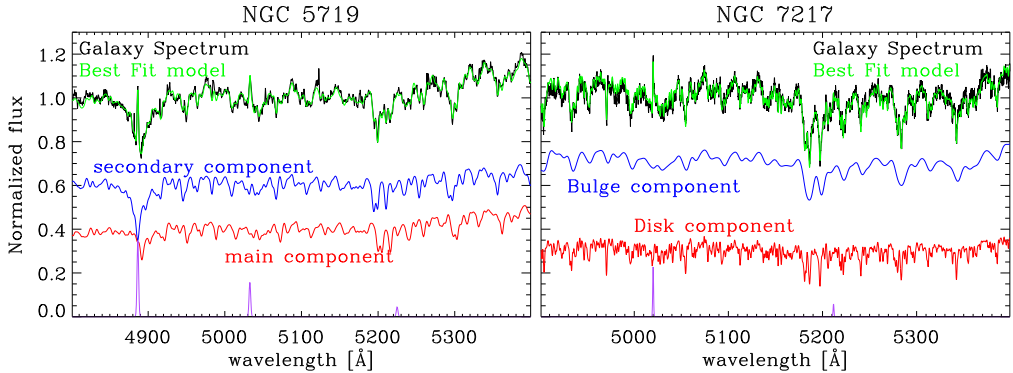} 
\caption{Example of the fit results of the spectral decomposition
  technique, applied to disentangle the contribution of two
  counter-rotating stellar disks (left panel, from
  \citealt{Coccato+11}), and the contribution of the stars in the
  bulge and in the disk (right panel, from
  \citealt{Fabricius+14}). Black: observed galaxy spectrum; green:
  best fit model; blue and red: spectra of the two stellar components;
  purple: gas emission lines.}\label{fig:decomposition}
\end{figure}

Figure \ref{fig:decomposition} shows some examples of results of the
spectral decomposition.

\section{Applications and results}

The technique has been successfully applied to disentangle (i)
counter-rotating stellar components in galaxies \citep{Coccato+11,
  Coccato+13, Pizzella+14}, (ii) the kinematics of stars in bulge and
disk \citep{Fabricius+14}, and (iii) the kinematics of the host galaxy
and orthogonal disk in polar disk galaxies \citep{Coccato+14}.

\begin{figure}
\centering
\hspace{-0.5cm}
\includegraphics[width=1.0\columnwidth]{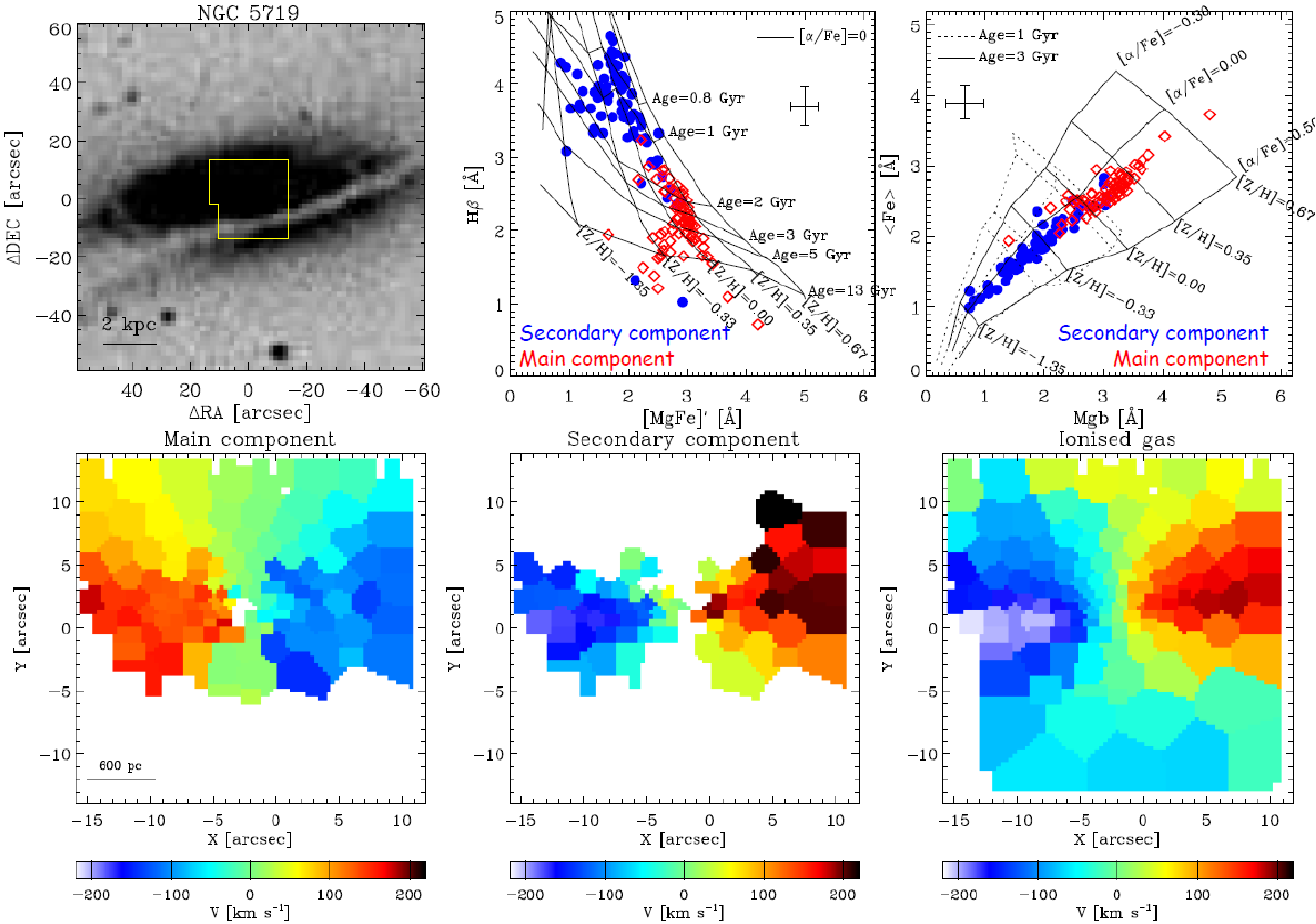} 
\caption{Top left panel: image of the Sa galaxy NGC 5719, the yellow
  region indicates the footprint of the VIMOS/IFU field of view. Top
  central and right panels: line strengths of the Lick indices
  measured on the main (red) and secondary (blue) stellar components:
  H$\beta$, combined magnesium-iron index [MgFe]' \citep{Gorgas+90},
  Mgb, and average iron index $<$Fe$>$ \citep{Thomas+03}. The black
  lines show the prediction of single stellar population models
  \citep{Thomas+11}. The bottom panels show the measured
  two-dimensional velocity fields of the main (left) and secondary
  (central) stellar components, and the ionized gas component
  (right). Adapted from \citet{Coccato+11}.}\label{fig:5719}
\end{figure}

Here, we describe the results obtained for a sample of galaxies that
host counter-rotating stellar disks of comparable size: NGC 3959, NGC
4183, NGC 4550, and NGC 5719. Observations were obtained with the
VIMOS integral field unit at the VLT (Chile), except for NGC 4138,
which was observed with the B\&C spectrograph at the 1.22-m telescope
at Padova University (Italy). In all the studied galaxies, we detect
the presence of an extended secondary stellar component, which is
counter-rotating with respect the more luminous and massive stellar
component. In addition, the secondary component is rotating along the
same direction as the ionized gas. The measurements of the equivalent
width of the absorption line features (H$\beta$, Mgb, Fe5270, and
Fe5335) reveal that the secondary stellar component is always younger
than the main stellar component, and it is more metal poor. Such a
difference in stellar population is particularly pronounced in NGC
5719 (Figure \ref{fig:5719}).

\begin{figure}
\centering
\includegraphics[width=1.0\columnwidth]{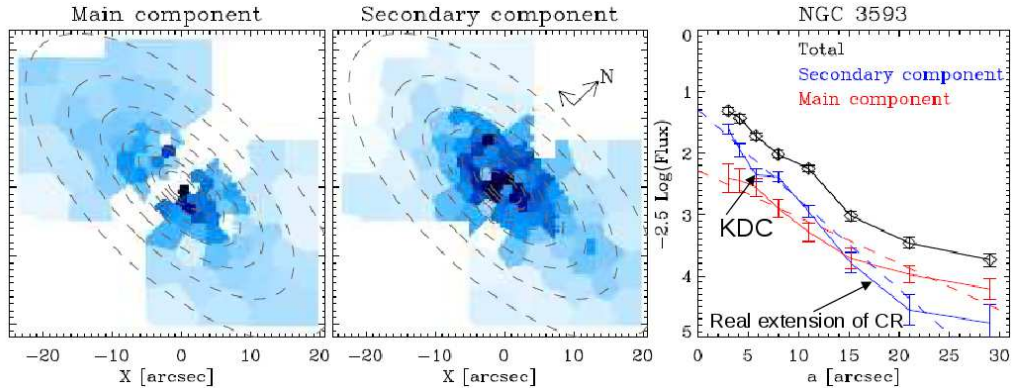} 
\caption{Maps (left and central panels) and radial profiles (right
  panel) of the surface brightness of the main and the secondary
  stellar components of NGC 3593. The black, red, and blue solid lines
  correspond to the total surface brightness and to the surface
  brightness of the main and the secondary stellar components,
  respectively. The dashed red and dashed blue lines correspond to the
  best-fitting exponential disks to the surface brightness of the main
  and secondary stellar components, respectively. Dashed ellipses
  represent the boundaries of the elliptical annuli where the median
  surface brightnesses were computed. Adapted from \citet{Coccato+13}.}\label{fig:surf_br}
\end{figure}

In the case of IFU observations, the spectral decomposition allows
also to investigate the morphology of the decoupled structures by
comparing their light contribution to the reconstructed image of the
galaxy. It is therefore possible to unveil the real extension and
geometrical properties such as orientation, ellipticity, and scale
length of the decoupled disks. In Figure \ref{fig:surf_br} we compare
the reconstructed images and the surface brightness radial profiles of
the counter-rotating disks observed in NGC 3593. The secondary
components overshines the main stellar disk within $\sim 10''$. A
single-component fit would reveal the presence of the secondary
component only in the inner regions, by revealing the kinematic
signature of a kinematically decoupled core (KDC,
e.g. \citealt{Bender88}).

The combination of the luminosity profile and stellar
mass-to-light ratio, as inferred from the stellar population analysis,
shows that the secondary component has a lower mass density radial
profile than the main component, also in the inner regions where it
dominates in light.

\section{Conclusions}

The stellar population properties and kinematics of the
counter-rotating stellar disks are consistent with the formation
scenario where a pre-existing galaxy acquired external gas on
retrograde orbits from a companion galaxy or the external medium
(e.g. \citealt{Thakar+96}; \citealt{Pizzella+04}). Depending on the
amount of acquired material, the new gas removes the pre-existing gas
and sets onto a disk in retrograde rotation. The new gas forms the
stars of the counter-rotating secondary component. In the case of NGC
5719 we have a direct of the on-going acquisition process
\citep{Vergani+07}.

Despite of their discovery more than 30 years ago, counter-rotating
galaxies still represent a challenging subject (see \citealt{Bertola+99, Corsini+14}
for reviews). We have showed that the spectral decomposition allows to
measure the properties of the stellar populations of both components,
and to probe the predictions of the different formation
scenarios. Forthcoming large integral field spectroscopic surveys such
as MANGA \citep{Bundy+14}, will allow to increase the statistics on
these objects, providing fundamental clues to constrain their
formation scenarios.


\end{document}